\documentclass[11pt, a4paper]{article}
\usepackage{jcappub}

\usepackage{natbib}

\usepackage{subfig}
\usepackage{amssymb, bm, graphicx, graphics, color, mathrsfs, hyperref}
\usepackage{geometry}
\newcommand{\be}{\begin{equation}}
\newcommand{\ee}{\end{equation}}
\newcommand{\ben}{\begin{eqnarray}}
\newcommand{\een}{\end{eqnarray}}

\newcommand{\la}{{\lambda}}

\newcommand{\cO}{{\cal O}}

\newcommand{\p}{\partial}
\newcommand{\na}{\nabla}

\newcommand{\tF}{\tilde F}

\newcommand{\ga}{\gamma}

\newcommand{\tA}{{\tilde A}}
\newcommand{\tB}{{\tilde B}}
\newcommand{\talpha}{{\tilde \alpha}}
\newcommand{\tzeta}{{\tilde \zeta}}

\graphicspath{{fig/}{./fig/}{.}}

\title{Holographic DC SQUID in the presence of {\it dark matter}}

\author[a]{Bart\l{}omiej Kiczek, }
\author[a]{Marek Rogatko}
\author[a]{and Karol I. Wysoki\'{n}ski }

\affiliation[a]{Institute of Physics \label{addr}\protect \\
Maria Curie-Sklodowska University \protect \\
20-031 Lublin, pl.~Marii Curie-Sk\l{}odowskiej 1, Poland}

\emailAdd{bkiczek@kft.umcs.lublin.pl}
\emailAdd{marek.rogatko@poczta.umcs.lublin.pl, rogat@kft.umcs.lublin.pl} 
\emailAdd{karol.wysokinski@umcs.pl}

\abstract{The gauge-gravity duality has been applied to examine the properties of holographic
superconducting quantum device (SQUID), composed of two S-N-S Josephson junctions,
 influenced by {\it dark sector} modelled by the additional $U(1)$-gauge field coupled to the ordinary Maxwell one. The {\it dark matter} sector is known to 
affect the properties of superconductors and is expected to enter the current-phase relation. The kinetic mixing between two gauge fields provides a mechanism allowing for the conceivable observation of the effect.
We find small but visible effect of the {\it dark matter} particle traversing the device, which shows up as a change of its maximal current.}

\keywords{Gauge-gravity correspondence, Holography and condensed matter physics(AdS/CMT), Black Holes, Dark Matter}

\arxivnumber{}

\begin{document}
\maketitle

\section{Introduction}
The quest for the {\it dark matter}  in the Universe has been one of the most important
topics of the current research in cosmology and physics \cite{ber18}.
Contemporary astronomical observations of galaxies and primordial radiation endorse the fact that our Universe is made mostly (over 23 percent of its mass) of non-luminous {\it dark matter}. Several new types of fundamental particles have been claimed  as candidates for {\it dark matter} sector. They are expected to interact with nuclei in suitable detector materials on Earth. It was claimed only by the DAMA collaboration \cite{ber98,ber13} that they observed modulation in the rate of interaction events which might be the trace of {\it dark matter} sector. Several groups want to reproduce the DAMA results but in vain \cite{cos18}. 
Recently the problem  of {\it dark matter} identity and its  detection has gained new impetus after the XENON1T experiment has reported the 
signal which source could not  be explained \cite{xen20}. This discovery has sparked intensive theoretical investigations 
of various scenarios which elucidate
the observed effect 
\cite{for20,bra20,bel20,bal20,tak20}. This situation triggers the discussion concerning the composition, interaction with ordinary matter, the self-interaction and the possible ways to discriminate between various models of {\it dark matter} \cite{masi2015}. 

It was argued that due to the growing sense of 'crisis' in the {\it dark matter} particle community, 
arising from the absence of evidences for the most popular candidates for 
{\it dark matter} particles like WIMPs, axions and sterile neutrinos, diversifying 
the experimental effort should be paid attention to. These efforts ought to accomplish
upcoming astronomical surveys, gravitational wave observatories which can provide 
us some complimentary information about the {\it dark matter} sector.
They constitute our best hope for making progress in this direction. 
Unconventional experiments and techniques are also being looked for. One of the prospect directions 
is related to clever usage of molecular or condensed matter systems 
including superconductors or superconducting devices. 
Recent proposals authorise search for bosonic {\it dark matter} {\it via} absorption 
in superconductors \cite{hochberg2016}, using superfluid helium \cite{knapen2017} or  
optical phonons in polar materials \cite{knapen2018} to detect light {\it dark matter}. More
exotic proposals are based on the observations of color centres production in crystals \cite{budnik2018}, 
or the usage of bulk three-dimensional Dirac semi-metals \cite{hochberg2018} and other topological 
semiconducting compounds \cite{liang2018}, as well as,
multilayered optical devices \cite{bar18}, for the detection of possible candidates for {\it dark matter} sector.

The theoretical approach we  have been using  in our research relies on the 
AdS/CFT correspondence \cite{mal98,wit98},
which has been found to provide strong coupling description \cite{sac12,zaanen-book} of many
condensed matter models. In particular this gauge-gravity duality has been shown to
describe $U(1)$ and $SU(2)$ symmetry breaking superconducting transitions.
The applications of the holography to study 
various models of superconductivity have been recently reviewed in \cite{cai-rev2015}.

Our previous AdS/CFT studies \cite{nak14}-\cite{rog18}
on holographic superconductors and superfluids reveal the influence 
of {\it dark matter} sector on various properties of them,
e.g., superconducting transition temperature, Ginzburg-Landau ratio, 
vortices and condensation flow, viscosity bound for anisotropic superfluids and
interacting currents in holographic Dirac fluid. These findings may in principle 
help to design future condensed matter experiments oriented on detection of
{\it dark matter}. However, more sensitive probes are still searched for. 
One of them is proposed in the current paper, i.e., examination of the
properties of DC SQUID device built of two Josephson junctions and influenced by {\it dark sector}.

The Josephson effect \cite{joseph63} is the well known quantum phenomenon observed in devices consisting of two superconductors separated by a thin layer of normal metal, insulator or a superconductor with much lower transition temperature. Such junctions, known as Josephson junctions, even without external bias sustain the direct current  $I$ which magnitude is related to the phase difference $\phi=\phi_L-\phi_R$
\be
I=I_{max}\sin\phi,
\label{eqjj}
\ee
where $\phi_L$ and $\phi_R$ are the phases of the superconducting order parameter on the left 
and right hand side of the junction. 
When the external magnetic flux $\Phi$ penetrates the interior of SQUID its maximal current depends on the flux as $I_{max}(\Phi)=I_c|\cos \frac{\pi \Phi}{\phi_0}|$, where $\phi_0=\frac{h}{2e}$is the magnetic flux quantum.
This relation is at the heart of many devices able to detect tiny magnetic fields \cite{tinkham}. 
As it has already alluded to, our aim is to study the effect of the {\it dark sector} particle on the holographic SQUID.
Recently the so-called second order Josephson effect has been proposed to detect 
axionic {\it dark matter} \cite{beck2013,wilczek2014}. 

The holographic model of Josephson junction was also paid attention to. In \cite{hor11} a gravitational 
dual of S-N-S junction was constructed and the calculations on the 
gravity side reproduced the standard relation between the current across the junction and the phase 
difference of the considered condensate. In the probe limit, Maxwell field
coupled to complex scalar one, was examined in the background of AdS-Schwarzschild black brane. The Josephson 
junction array was considered in \cite{kir11}, where a model was proposed in the background of multi-supergravity
theories on products of distinct asymptotically AdS spacetimes coupled by mixed boundary conditions. Among all it was found that the Cooper-pair condensates were described by a discretised Schr\"odinger-like equation. In a continuum limit the relation in question became a generalised Gross-Pitajevski equation, known
from the long-wavelength description of superfluids.

A holographic model of S-I-S junction constructed by examining a complex scalar field coupled to Maxwell one, in the background of four-dimensional
AdS soliton was examined in \cite{wan12}. The dependence of the maximal current on the dimension of the condensate operator and the width of the junction were
presented. 
On the other hand, a holographic configuration including a chiral time-reversal breaking superconductors in $(2+1)$-dimensions was discussed in \cite{roz14},
where the ansatz for $p_x+ip_y$ superconductors \cite{rob08} was implemented. Such kind of superconductor is believed to support topologically protected gapless
Majorana-Weyl edge modes. 
On the other hand, the model with two coupled vector fields, was also implemented in a generalisation of p-wave superconductivity, for the holographic model of ferromagnetic 
superconductivity \cite{amo14}.

The construction of a Josephson junction in non-relativistic case with a Lifshitz geometry as the dual gravity was discussed in \cite{li14}, where the effect of the
Lifshitz scaling was elaborated. Among all the standard sinusoidal relation between the current and phase difference was revealed for various values of scaling. 
The relation featuring 
exponential decreasing  between condensate operator and the width of the weak link, as well as, the relation connected the critical current with the width were found. 

Holographic s-wave and p-wave Josephson junctions with backreaction were studied in \cite{wan16}. It turns out that the critical temperature of the considered 
junctions decreases with the increase of backreaction, while the tunnelling current and condensation also decrease with the growth of backreaction. However,
the relation between current and phase difference remains in the form of a sine-function.
On the other hand, the problems of holographic models of hybrid and coexisting both types of junctions, i.e., s- and p-wave were analysed in \cite{liu15}.

The studies of S-N-S junction in massive gravity theory unveil that the graviton mass parameter make it difficult for the normal metal-superconductor phase transition to occur
\cite{hu16}. Moreover the mass graviton parameter increase will cause the decrease of the maximal tunnelling current.

\begin{figure}
\centering
\includegraphics[width=0.5\linewidth]{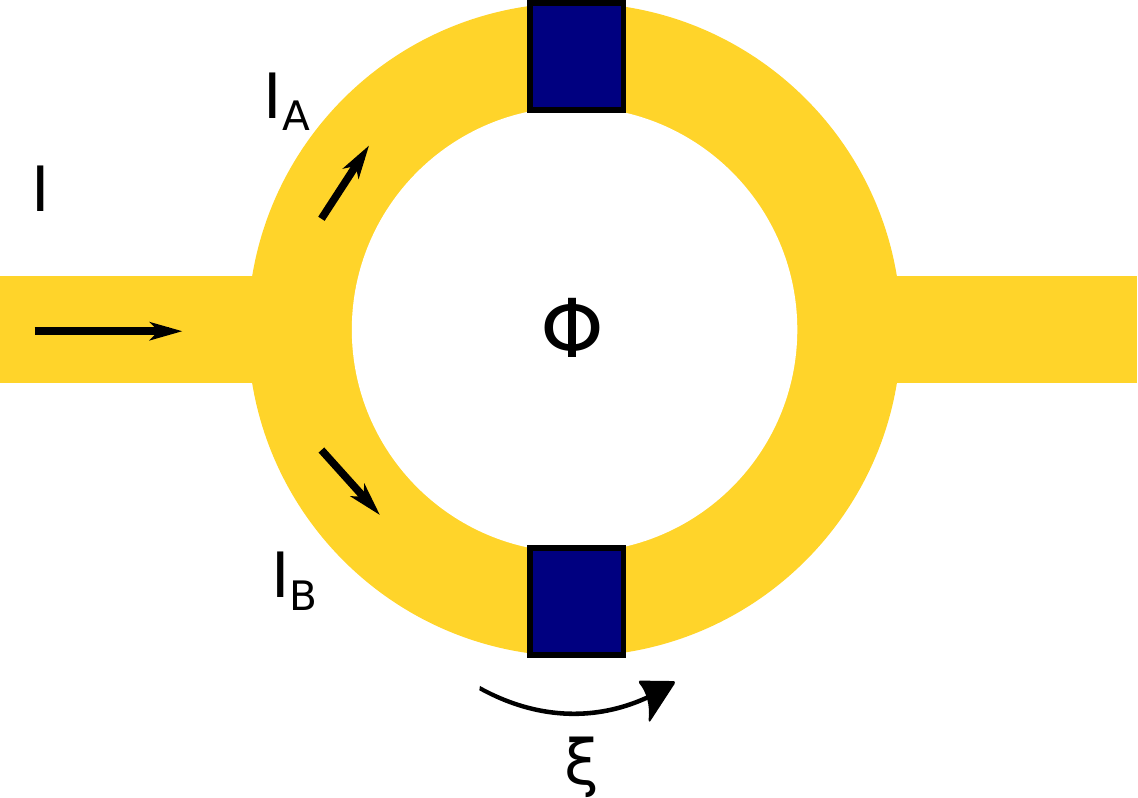}
\caption{The schematic view of the condensed matter superconducting quantum interference device consisting of two Josephson junctions (dark-blue) embedded into the superconducting loop (yellow-bright). The current $I$ flowing across the system is split into $I_A$ in the upper and $I_B$ in the lower branch of the device. The presence of the magnetic flux $\Phi$ induces the interference in the system. For the holographic setup see text for details.}
\label{fig2}
\end{figure}

In recent years there has been also an interest in building holographic model of superconducting quantum interference device (SQUID) consisting of two appropriately 
connected Josephson junctions (see figure \ref{fig2} for the condensed matter model).
Consequently, in order to construct the holographic analog of SQUID one considers Einstein-Maxwell gravity with complex scalar field in $(3+1)$-dimensional AdS Schwarzschild black brane background.
One of the spatial dimensions is compactified into a circle and with the properly chosen profiles of the chemical potential constitute the main building 
blocks of the holographic SQUID \cite{caisq14,tak15}.

Taking our main goal into account we interpret the relation $I=I_A+I_B$ for the total current in the SQUID, with  the currents flowing across each of the junctions given by 
the relation
 (\ref{eqjj}), in the slightly different manner.  Namely suppose, that the 
phase difference,  $\Delta \phi_A$ (or $\Delta \phi_B$), in one of the  branches of the considered device has changed as a result of external factors. 
In the model under inspection we understand this phenomenon as 
a {\it hidden sector} particle appearing in one of the Josephson junctions of the SQUID. 
This means that the current phase relation changes in the same way as if the 
effective magnetic flux would appear. 
The large sensitivity of the device, rises the hope of detecting such transition of the {\it dark matter} particle, under the assumption of the validity
of the coupling between {\it hidden} and {\it visible} sectors.
 In this paper we shall suppose that a single {\it dark 
matter} particle appears in one SQUID branch at a time, leaving aside the issue of possible fluctuations 
of the SQUID signal due to consecutive events. This problem will be studied in future publications.

The main motivation standing behind our consideration is to reinvestigate the holographic model of the Josephson junctions and the SQUID
\cite{hor11,caisq14,tak15},
in order to include the {\it hidden sector} particles responsible for the {\it dark matter} fields. In conventional physics the {\it hidden sector} should be added to the Standard Model. 
On the contrary, in our attitude we shall build holographic model in which {\it dark sector} naturally emerges from the string/M theory.
In our research we shall consider the model in which the additional $U(1)$-gauge field is coupled to the ordinary Maxwell one. 
The action describing Maxwell {\it dark matter} sector is provided by \cite{vac91,ach00}.
\be
S_{dm + EM} = \int \sqrt{-g} ~d^4 x  \big( 
- \frac{1}{4} F_{\mu \nu} F^{\mu \nu} 
- \frac{1}{4}B_{\mu \nu} B^{\mu \nu}
 - \frac{\alpha}{4} F_{\mu \nu} B^{\mu \nu} \big),
\label{sgrav} 
\ee
where $F_{\mu \nu} = 2 \nabla_{[ \mu} A_{\nu ]}$ stands for the ordinary Maxwell strength tensor, while
the hidden $U(1)$-gauge field $B_{\mu \nu}$ is given by $B_{\mu \nu} = 2 \nabla_{[ \mu} B_{\nu ]}$. $\alpha$ is a coupling constant between two gauge fields.
Predicted values of $\alpha$-coupling constant, being
the kinetic mixing parameter between the two $U(1)$-gauge fields, for realistic string compactifications range between $10^{-2}$ and $10^{-16}$ \cite{abe04}-\cite{ban17}.

The compatibility with the current observations authorises its order up to $10^{-3}$.
It happens that some astrophysical phenomena like observations of $511$ eV gamma rays
\cite{integral}, experiments connected with the detection of the electron positron excess in galaxies
\cite{atic,pamela}, as well as, possible explanation of muon anomalous magnetic moment \cite{muon}, 
strongly argue for the idea of {\it dark sector} coupled to the Maxwell one.
Moreover, the {\it kinetic mixing term} between ordinary boson and relatively light one (the {\it dark} one) arising from $U(1)$-gauge symmetry bounded with a {\it hidden sector},
may be responsible for a low energy parity violation \cite{dav12}. On the other hand, the low energy gauge interaction in the {\it hidden sector} 
may be envisaged by the Higgs boson decays,
and a relatively light vector boson with mass $m\ge 10GeV$ emergence \cite{dav13}.

As was mentioned the justifications of such kind of models might be obtained from the top-down perspective \cite{ach16}, 
starting from the string/M-theory. It is very important from the AdS/CFT correspondence point 
of view, because string/M-theory constitutes a fully quantum
description and this guarantees that any phenomena envisaged 
by the top-down reduction are physical.

The second term in the above action is connected with some hidden sector \cite{ach16}, 
while the interaction of the visible and hidden sectors are
described by the so-called {\it kinetic mixing term}, for the first time introduced 
in  \cite{hol86}, in order to describe the existence and subsequent integrating out 
of heavy bi-fundamental fields charged under the $U(1)$-gauge groups. The term in question 
is characteristic for theories having in addition to some visible gauge group an
additional one, in the hidden sector. The hidden sector describes states in the low-energy 
effective theory, uncharged under the the Standard 
Model gauge symmetry groups. On the contrary they are charged under their own groups. 

It is gravity which enables interaction between those two sectors \cite{portal1,portal2}. 
It turns out that the realistic embeddings of the Standard Model in $E8 \times E8$ string theory, 
as well as, in type I, IIA, or IIB open string theory with branes, require the existence 
of the hidden sectors for the consistency and supersymmetry breaking \cite{abe08,die97}. 
On the other hand,  in string phenomenology \cite{abe08} the dimensionless kinetic mixing term 
parameter can be produced at an arbitrary high energy scale and it does not deteriorate from 
any kind of mass suppression from the messenger introducing it. This fact is of a great importance 
from the experimental point of view, due to the fact
that its measurement can provide some interesting features of 
high energy physics beyond the range of the contemporary colliders.

Moreover the mixing term of two gauge sectors are typical for states for open 
string theories, where both $U(1)$-gauge groups are advocated by D-branes that are separated in extra dimensions. It happens
in supersymmetric Type I, Type IIA, Type IIB models and results in the existence of massive open strings 
which stretch between two D-branes \cite{abe04}.

The organization of the paper is as follows. In the next section we pay attention to the holographic model of 
DC SQUID composed of two 
Josephson junctions in Einstein-Maxwell gravity
with complex scalar field affected by the {\it dark sector} particles modelled by the hidden $U(1)$-gauge field interacting with the ordinary Maxwell one.
One considers the model of the junction,  where a central part constitutes  a normal metal (S-N-S). We also mentioned the case with an insulator (S-I-S).  
Sec. 3 is devoted to the 
numerical solutions of the adequate equations of motion, for the homogeneous case and the case of a {\it dark matter} beam.
We propose an 'experiment' which helps to distinguish the presence of
{\it dark matter} sector. In Sec. 4 we concluded our investigations.

\label{sec:intro}
\section{The holographic model and equations of motion}
\label{sec:model}
In this section we present the examined holographic model and the resulting equations of motion. We shall examine
the holographic DC SQUID, composed of the two S-N-S Josephson junctions. We give some comments concerning the construction of holographic model of such a device.

\subsection{How to build holographic SQUID influenced by {\it dark matter} sector?}
As the SQUID is composed of two Josephson junctions we shall concentrate on the problem how to build holographic model of the S-N-S Josephson junction with the phase difference across the normal region, say $\ga$.
In order to simulate it one has to implement the holographic description of the superconductor, i.e., we should have the AdS background because of the fact that we describe 
an approximately conformal field theory. 

In the next step we want to 
describe the conserved $U(1)$ currents  $J_\mu$ in the field theory. According to the AdS/CFT dictionary\cite{{zaanen-book}} in the gravity dual they correspond to gauge fields. In our case these gauge fields will be bounded with ordinary Maxwell field and the auxiliary one, pertaining to the {\it hidden sector}.

The superconducting state requires spontaneous symmetry breaking\footnote{The global symmetry of the boundary field theory is the conservation of the charge, which is precisely broken below the superconducting transition temperature. On the gravity side this is marked as an instability of the black hole which becomes hairy.}, which on the gravity side corresponds to the condensation of the (complex) scalar field. The operator for which vacuum expectation value is non-zero will correspond to the scalar field in the bulk.

In order to incorporate temperature in the theory we take into account  
the black brane with Hawking temperature. On the other hand, to have $T_C \ne 0$, one incorporates the scale to generate it, which can be 
done by either introducing  chemical potential $\mu$ or a non-zero charge density. In the AdS/CFT dictionary $\mu$, on the field theory side, is correlated with the vacuum expectation value of the time component of gauge field. Thus if we take a non-zero chemical potential to generate superconductivity, one achieves
a critical temperature proportional to it. All these points lead the holographic model of Josephson junction in which one should also have varying chemical potential $\mu (\vec x)$ at fixed temperature 
(we have black brane at fixed Hawking temperature)
such that in most of the space temperature $T$ is below $T_C$ connected with $\mu$ (the system in question is in a superconducting state) but
in a narrow gap is above it,  and the system is in a normal state.

For technical reasons we introduce gauge invariant fields described by $M_\mu = A_\mu - \p_\mu \phi$, which have the adequate asymptotics at the boundary of AdS spacetime,
$i.e.$ for $r \rightarrow \infty$ proportional to
chemical potential, charge density, superfluid velocity and current. Because of the fact that the ansatz for the scalar field has a quantum phase $\phi$ and
$ M_x = - \p_x \psi + A_x$, the gauge invariant phase difference across the normal gap is given by
\be
\ga = - \int_{gap} dx M_x = \Delta \phi - \int_{gap} dx A_x.
\ee
It concludes, that from the boundary at infinity one obtains the following relation:
\be
\ga = - \int_{- \infty}^{+ \infty} dx \Big[ \nu(x) - \nu(\infty) \Big].
\ee

Moreover, at the black brane event horizon we need to have the regularity conditions fulfilled, i.e., $M_t (r = r_0) = 0$. One also imposes that the junction is unbiased so $\mu(\infty) = \mu(-\infty)$ meaning that  
near the boundary $\mu$ approaches  $x$-independent value.

Finally, in order to construct the holographic model of Josephson junction one demands the adequate profile for the chemical potential, for which we ought to obtain 
that $T_C \sim \mu(\infty)$ with $T_{Hawking} < T_C$, and also we demand that $\mu$ is much smaller in the junction leading to $T_C<T_{Hawking}$ there.
All these conditions ensure 
that we arrive at superconductor for most of the range of $\vec x$, only in the thin region we shall get a normal state.

To conclude, the philosophy underlying our considerations is to incorporate {\it visible} and {\it hidden} sectors into AdS spacetime, apply the AdS/CFT machinery, and look
if there are some implications of the auxiliary $U(1)$-gauge field at $r \rightarrow \infty$, at the boundary of the spacetime.

\subsection{Equations of motion for the model}

The holographic
model of the 2+1 dimensional system in question relies on the 3+1 dimensional gravitation 
background of AdS static Schwarzschild black brane spacetime. One will work in the probe limit, which means that matter field do not cause the fluctuations of
the gravitational one, we have no backreaction effects. In the probe limit approximation the dynamics of the gravitational part of the considered model will
decouple from the rest, therefore one can elaborate gauge field perturbations on the fixed black brane background which is therefore uncharged.

The gravitational action in (3+1) dimensions is taken in the form
\be
S_{g} = \int \sqrt{-g}~ d^4 x~  \bigg( R - \Lambda\bigg), 
\label{graction}
\ee
where $\Lambda = - 6/ L^2$ stands for the negative cosmological
constant, while $L$ is the radius of the AdS spacetime. In what follows we set $L=1$.

We shall examine 
the Abelian-Higgs sector coupled to the {\it hidden} $U(1)$-gauge field, coupled to the {\it visible } one by the {\it kinetic mixing term}, with a coupling $\alpha$
\ben \nonumber
\label{s_matter}
S_{dm+EM+H} = \int \sqrt{-g}~ d^4x  \bigg( 
&-& \frac{1}{4}F_{\mu \nu} F^{\mu \nu} - \left [ \nabla_{\mu} \psi - 
i ~A_{\mu} \psi \right ]^{\dagger} \left [ \nabla^{\mu} \psi - i~  A^{\mu} \psi  \right ]
- m^2 |\psi|^2 \\ 
&-& \frac{1}{4} B_{\mu \nu} B^{\mu \nu} - \frac{\alpha}{4} F_{\mu \nu} B^{\mu \nu}
\bigg).
\een  
In many analysis of the {\it hidden sector} in holographic approach \cite{nak14}-\cite{rog18}, 
$\alpha$ is taken as a constant. In \cite{amo14}, where the generalisation of the considered model to $SU(2)$ gauge fields was studied in the context of holographic
model of ferromagnetic superconductivity, $\alpha$ was incorporated in the redefined gauge fields being the mixture of {\it visible} and {\it hidden} sectors components.
In our attitude we are interested in the influence of the {\it dark sector} on the physics and therefore we shall treat both sectors and kinetic mixing term between them.

However, as it will be revealed in section 3.1 of our work, such assumption about the homogeneity of $\alpha$-coupling constant gives no response on the considered
holographic model of SQUID. 

In order to make the investigations more realistic, and find the influence of
{\it dark sector} on the physics of the holographic device, we suppose that 
instead of 
the coupling in the {\it kinetic mixing term} we shall take into account the strength parameter function $\zeta$, which controls the strength of {\it visible} and
{\it hidden sectors} interaction. Having in mind that our line element is static, i.e., that one has time-like Killing vector field 
orthogonal to the three-dimensional space-like hypersurface, our strength parameter function will depend on spatial coordinates.

In our considerations we shall elaborate the problem of the
response of a holographic superconducting junction/ SQUID to the appearance 
of {\it dark matter particle}. Due to the geometric constraints (see for the details \cite{caisq14,tak15}), the nontrivial response of the holographic (2+1) dimensional SQUID  is expected to appear only  if one of its junctions is affected by a {\it dark matter particle}. 

As it was already mentioned,  
to mimic the passage of the {\it dark particle} {\it via} one of the junctions of the holographic SQUID we 
assume that  the coupling $\zeta$ depends on space-like coordinates. This is taken into account 
in the derivation of the underlying equations of motion. Later on we specify the particular dependence $\zeta(x_a)$ used 
to solve the model.

The equations of motion are provided by
\ben \label{e1}
\bigg( \na_\mu &-& i~A_\mu \bigg) \bigg( \na^\mu - i ~A^\mu \bigg)\psi - m^2~\psi = 0,\\ \label{e2}
\na_\mu F^{\mu \nu} &+& \frac{1}{2} \na_\mu \Big( \zeta ~B^{\mu \nu} \Big) - i ~\bigg[ \psi^\dagger~
\bigg( \na^\nu - i ~A^\nu \bigg)~ \psi - \psi~\bigg( \na^\nu + i ~A^\nu \bigg) \psi^\dagger \bigg] = 0,\\ \label{eqb}
 \na_\mu B^{\mu \nu} &+& \frac{1}{2} \na_\mu \Big( \zeta ~F^{\mu \nu } \Big)= 0.
\een
Consequently, using the last two equations one has that
\be
\tilde \zeta~\na_\mu F^{\mu \nu}  + \frac{1}{2} \na_\mu \zeta \Big(
B^{\mu \nu} - \frac{1}{2} \alpha F^{\mu \nu} \Big) - i ~\bigg[ \psi^\dagger~
\bigg( \na^\nu - i ~A^\nu \bigg)~ \psi - \psi~\bigg( \na^\nu + i ~A^\nu \bigg) \psi^\dagger \bigg] = 0,
\ee
where we set $\tilde \zeta = 1- \zeta^2/4$.

As we have mentioned above, as a gravitational background we shall consider static four-dimensional AdS-Schwarzschild black brane line element, given by
\be
ds^2 = - f(r) dt^2 + \frac{1}{f(r)}dr^2 + r^2~(d \xi^2 + dx^2),
\label{adsschw}
\ee
where $f(r) = r^2 -r_0^3/r$ and $r_0$ is the horizon radius of the black brane, while the direction $\xi$ is compactified with the periodicity $- \pi~R \le \xi \le \pi~R$, 
where $R$ is the radius of the $\xi$-loop.

For further analysis it is convenient to assume that $r_0 = 1$ and $\pi~R = 10$.
According to the claim in \cite{hor11}, the coherence length for the considered model is 
 estimated to be about $\sim 1.20$ within these units. 
We argue that the length of the SQUID loop should be much larger than the coherence length, 
in order to decrease the influence of the proximity effect and  consequently the mixing of the phases. 
Fulfilling this condition, we obtain a homogeneous superconducting phase at the both 
ends of the $\xi$-direction. It results in a well defined phase difference for the junction under 
inspection. On the other hand, one wants to establish our weak link of the junction to be narrow enough, that its critical current receives a sufficiently high value. This is important from the point of view of the numerics, i.e., the precision of the solutions. Having all the above arguments in mind, we establish $R = 10/\pi$ as a reasonable choice.

In what follows one chooses the following components of the fields in the underlying theory:
\be
\psi = \vert \psi \vert e^{i \phi}, \qquad
A_{\mu} = (A_t,~A_r,~A_{\xi},~0), \qquad B_{\mu} = (B_t,~B_r,~B_{\xi},~0),
\ee
where we assume that all the field components and the phase are real functions dependent on $(r,~\xi)$-coordinates. 
Moreover, one defines the gauge invariant  quantities $M_\mu = A_\mu - \p_\mu \phi.$ 

Consequently, the equations of motion yield
\ben
\p^2_r \vert \psi \vert &+& \frac{1}{r^2~f} \p^2_{\xi} \vert \psi \vert  + \bigg( \frac{2}{r} +
 \frac{\p_r f}{f} \bigg)~\p_r \vert \psi \vert  + 
\bigg[\frac{1}{f^2}~M_t^2~ 
- \frac{1}{r^2~f}~M_{\xi} - M_r^2
\\ \nonumber
&-&  \frac{m^2 ~}{f}\bigg]\vert \psi \vert  = 0,\\
\p_r M_r &+& \frac{1}{r^2 f}\p_{\xi} M_{\xi} + \frac{2}{\vert \psi \vert } \bigg(
M_r \p_r\vert \psi \vert  + \frac{M_{\xi}}{r^2 f}~\p_{\xi} \vert \psi \vert  \bigg)  + \bigg( \frac{2}{r} + \frac{\p_r f}{f}  \bigg)M_r = 0,\label{eq:constraint}\\
\tzeta \Big[ \p^2_r M_t &+& \frac{2}{r}~\p_r M_t + \frac{1}{r^2~f}~\p^2_{\xi} M_t \Big] - \frac{2}{f}~M_t~\vert \psi \vert ^2  \\ \nonumber
&+& \frac{1}{2} \p_r \zeta \Big[
\Big( \p_t B_r - \p_r B_t \Big) - \frac{\zeta}{2} \Big( \p_t M_r - \p_r M_t\Big) \Big] \\ \nonumber
&+& \frac{1}{2} \p_{\xi} \zeta \Big[ - \frac{1}{r^2 f} \Big(\p_{\xi}B_t - \p_t B_{\xi} \Big) + \frac{\zeta}{2} \frac{1}{r^2 f} \Big( \p_{\xi} M_t - \p_t M_{\xi} \Big) \Big] =0,\\
\tzeta \Big[
\p^2_{\xi} M_r &-& \p_{\xi} \p_r M_{\xi} \Big] - 2~r^2~M_r~\vert \psi \vert ^2  + \frac{1}{2} \p_{\xi} \zeta \Big[
\Big( \p_{\xi} B_r - \p_r B_{\xi} \Big)  \\ \nonumber
&-& \frac{\zeta}{2} \Big( \p_{\xi} M_r - \p_r M_{\xi} \Big) \Big] =0,\\
\tzeta \Big[
\p^2_r M_{\xi} &-& \p_r \p_{\xi} M_r + \frac{\p_r f}{f}~\bigg( \p_r M_{\xi} - \p_{\xi} M_r \bigg) \Big] - \frac{2~M_{\xi}}{f}~\vert \psi \vert ^2 \\ \nonumber
&+& \frac{1}{2} \p_r \zeta \Big[ \frac{f}{r^2} \Big( \p_r B_{\xi} - \p_{\xi} B_r \Big)
- \frac{\zeta}{2} \frac{f}{r^2} \Big( \p_r M_{\xi} - \p_{\xi} M_r \Big) \Big] =0.
\een
The components of the relation (\ref{eqb}) imply
\ben
\frac{2}{r} \p_r B_t  &+& \p_r^2 B_t + \frac{1 }{r^2 f} \p_{\xi}^2 B_t  + \frac{1}{2} \p_r \zeta~ \p_r M_t  + \frac{1}{2} \p_{\xi} \zeta \frac{1}{r^2 f} \p_{\xi} M_t \\ \nonumber
&+& \frac{\zeta}{2} \Big(
\frac{2}{r} \p_r M_t  + \p_r^2 M_t + \frac{1 }{r^2 f} \p_{\xi}^2 M_t \Big) = 0,
\label{eq2_a} \\
\p_r^2 B_{\xi} &-& \p_r \p_{\xi} B_r + \frac{\p_r f}{f} \Big( \p_r B_{\xi} - \p_{\xi} B_r \Big) + \frac{1}{2} \p_r \zeta \Big( \p_r M_{\xi} - \p_{\xi} M_r \Big) \\ \nonumber
&+& \frac{\zeta}{2} \Big[
\p_r^2 M_{\xi} - \p_r \p_{\xi} M_r + \frac{\p_r f}{f} \Big( \p_r M_{\xi} - \p_{\xi} M_r \Big) \Big] = 0,
\label{eq2_b} \\
\p_{\xi} \Big( \p_{\xi} B_r &-& \p_r B_{\xi} \Big) + \frac{1}{2} \p_{\xi} \zeta \Big( \p_{\xi} M_r - \p_r M_{\xi} \Big) +
\frac{\zeta}{2} \p_{\xi} \Big( \p_{\xi} M_r - \p_r M_{\xi} \Big) = 0.
\label{eq2_c} 
\een

To proceed further, we ought to choose the form of $\zeta(x_i)$. One of  the possible representations of it
is a Gaussian function form
\be
\zeta(\xi)=\alpha_0 e^{-(\xi-\xi_0)^2/\lambda^2},
\label{eqn:gauss_alpha}
\ee  
where $\alpha_0$ is the peak value of the coupling and $\lambda$ its decay length (half of the Gaussian width). 
Our choice of the strength function controlling interaction of both sectors in question $\zeta(\xi)$, was motivated by the features of static line element chosen for the 
description of holographic device, given by (\ref{adsschw}), as well as, the astrophysical observations revealing the space-like dependence of {\it dark matter} distribution
in the observable Universe.

The half of the Gaussian width $\lambda$ is an important parameter of the model as its ratio  to the holographic SQUID characteristic length-scale $\pi R$  and the shape of the weak link affect the holographic SQUID's response.

In principle, there are two possible attitudes for modelling the local presence of the {\it dark matter sector} fields in this 
theory. We can either pick $\alpha$ to be constant and 
impose specific boundary conditions for {\it dark matter} fields or as it was mentioned before, promote 
the strength function $\zeta$ with a spatial dependence and keep the fields in the homogeneous scenario.
In this approach we selected the latter one, so we can choose the $x$-independent and source free solution of $B_{\mu}$ fields, which simply yields 
\be
B_{\alpha} ~dx^{\alpha} = \mu_D \left(1 - \frac{r_0}{r} \right) dt.
\ee
By putting this assumption into our system of differential equations we receive a significant simplification, with only four functions to be obtained numerically. In particular the resulting equations depend 
on the {\it dark sector} only {\it via} the effective coupling $\tzeta$ and not $\mu_D$.
\ben \nonumber
\p^2_r \vert \psi \vert &+& \frac{1}{r^2~f} \p^2_{\xi} \vert \psi \vert  + \bigg( \frac{2}{r} +
 \frac{\p_r f}{f} \bigg)~\p_r \vert \psi \vert  + 
\bigg[\frac{1}{f^2}~M_t^2~ \label{eq:systemstart}
- \frac{1}{r^2~f}~M_{\xi}^2 - M_r^2 
\\ 
&-&  \frac{m^2 ~}{f}\bigg]\vert \psi \vert  = 0,\\
\tzeta \Big[ \p^2_r M_t &+& \frac{2}{r}~\p_r M_t + \frac{1}{r^2~f}~\p^2_{\xi} M_t \Big] - \frac{2}{f}~M_t~\vert \psi \vert ^2  \\ \nonumber
&+& \frac{1}{2} \p_{\xi} \zeta \frac{\zeta}{2} \frac{1}{r^2 f} \p_{\xi} M_t =0,\\
\tzeta \Big[
\p^2_{\xi} M_r &-& \p_{\xi} \p_r M_{\xi} \Big] - 2~r^2~M_r~\vert \psi \vert ^2  - \frac{1}{2} \p_{\xi} \zeta\frac{\zeta}{2} \Big( \p_{\xi} M_r - \p_r M_{\xi} \Big)  =0,\\
\tzeta \Big[
\p^2_r M_{\xi} &-& \p_r \p_{\xi} M_r + \frac{\p_r f}{f}~\bigg( \p_r M_{\xi} - \p_{\xi} M_r \bigg) \Big] - \frac{2~M_{\xi}}{f}~\vert \psi \vert ^2  =0. \label{eq:systemend}
\een
The above forms of the equations envisage that comparing to the case when $\zeta=0$, the modification is significant. Some terms
are multiplied by $\tzeta$ and moreover we have additional terms with $\zeta \p_\xi \zeta$ factor. Numerical solutions of these equations will be
presented in the subsequent part of the paper.

As we have remarked at the beginning of the section, the presented model describes SQUID with S-N-S type Josephson junctions.
Nevertheless, for the completeness of the results, we shall mention the modeling of S-I-S Josephson junction.
 In the holographic model of the insulator the AdS soliton line element will play the crucial role.
Namely, performing the double Wick rotation on the metric (\ref{adsschw}), we arrive at the 
AdS soliton line element \cite{hor98}. It is provided by
\be
ds^2 = - r^2 dt^2 +  \frac{dr^2}{f(r)} + r^2 dx^2 + f(r) d \chi^2,
\ee
with the same function $f(r)$ as before. However, the coordinate $\chi$ has the period 
$\beta = 4 \pi l/ 3r_0$, in order to avoid the conical singularity at $r = r_0$. 

As in the previous case we assume that the following components of the gauge field are non-zero 
$ A_{\mu} = (A_t,~A_r,~A_x,~0)$,
and that all the field components and the phase $\phi$ are real functions dependent on $(r,~x)$-coordinates.
The equations of motion are very similar to the ones obtained in the previous case. 
in fact only the factor in front of $\partial_r M_t$ instead of $\frac{2}{r}$ reads $\frac{\p_r f}{f}$.  
The results will differ only quantitatively, therefore they will be not analysed
numerically.

Having discussed the model set up we shall describe appropriate boundary conditions for our model.
Starting with AdS boundary where $r \rightarrow  \infty$, the asymptotic of the fields in question are provided by
\ben
\vert \psi \vert &=& \frac{\vert \psi^{(1)}(\xi) \vert}{r^{\Delta_{(1)}}} +  \frac{\vert \psi^{(2)}(\xi) \vert}{r^{\Delta_{(2)}}}
 + \cO(r^{-3}),\\
M_t &=& \mu(\xi) - \frac{\rho(\xi)}{r} + \cO(r^{-2}),\\
M_r &=& \cO(r^{-3}),\\
M_\xi &=& \nu(\xi) + \frac{J(\xi)}{r} + \cO(r^{-2}),
\een
where 
\be
\Delta_{(1)} = \frac{1}{2}~(3 - \sqrt{9 + 4m^2}), \qquad \Delta_{(2)} = \frac{1}{2}~(3 + \sqrt{9 + 4m^2}).
\ee
In the boundary field theory the quantities $\mu(\xi),~\rho(\xi),~\nu(\xi), $ and $J(\xi)$ are connected with the chemical potential,
charge density, superfluid velocity and current, respectively. On the other hand, 
$\vert \psi^{(1)}(\xi) \vert$ and $\vert \psi^{(2)}(\xi) \vert$
may be considered as the source and the vacuum expectation value of the dual operator of the scalar field $\vert \psi(\xi)^{(2)} \vert = \langle O \rangle$. In the numerical 
calculation we set
$\vert \psi^{(1)}(\xi) \vert = 0$ (as turning off the source), because one has that the $U(1)$-gauge symmetry be broken.
At  the black brane event horizon we demand that $g^{tt}~M_t$ ought to be regular, but because 
of the fact that $g^{tt}$ is divergent on the horizon, $M_t$ should vanish at $r=r_0$.
The remaining fields should be finite at the event horizon and the simplified equations of motion place the boundary conditions therein.
In our numerical calculations for simplicity we set $m^2 = -2$ which is above Breitenlohner-Freedman limit $m^2 \ge -\frac{9}{4}$, 
therefore $\Delta_{(1)} = 1$ and $\Delta_{(2)} = 2$.

\section{Numerical results}
This section will be devoted to 
the results of our numerical solutions of the equations of motion for the holographic Josephson junction and holographic DC SQUID device.
At first we pay attention to the case of homogeneous holographic superconductor influenced by the {\it dark matter} sector.
Then we proceed to analyse the 
model of SQUID with a beam of {\it dark matter}, i.e., we 
take into account the strength parameter controlling the strength of {\it visible} and {hidden} sectors interaction and allow for its
$\xi$ dependence.

\subsection{Homogeneous superconductor - warm up} 
\label{sec:homogeneous}
To commence with, let us consider the properties of homogeneous holographic superconductor, i.e., we take $\alpha$ as a constant value,  
influenced by the {\it dark matter} sector, and look for its
critical temperature, critical chemical potential and value of the condensation.
For this calculation we straight out the $\xi$ dimension renaming it to $x$.
Then by neglecting the spatial dependence of $x$-coordinate in the equations of motion, we receive 
the simplified system of differential equations
\ben
\p^2_r \vert \psi \vert &+& \bigg( \frac{2}{r} +
 \frac{\p_r f}{f} \bigg)~\p_r \vert \psi \vert  + 
\bigg[\frac{1}{f^2}~M_t^2~ 
- \frac{1}{r^2~f}~M_x - \frac{m^2 ~}{f}\bigg]\vert \psi \vert  = 0, 
\label{eq:hom1} \\
\p^2_r M_t &+&  \frac{2}{r} \p_r M_t - \frac{2}{\tilde \alpha~f}~M_t~\vert \psi \vert ^2 = 0, \label{eq:hom2} \\
\p^2_r M_x &+& \frac{\p_r f}{f}\p_r M_x - \frac{2~M_x}{\tilde \alpha~f}~\vert \psi \vert ^2 = 0.
\label{eq:hom3}
\een
where $\talpha = 1 -\frac{\alpha^2}{4}$ and all the above three functions, i.e., $\vert \psi \vert,~M_t,~M_x$, depend only on $r$.

\begin{figure}
\centering
\subfloat{\includegraphics[width=0.5\linewidth]{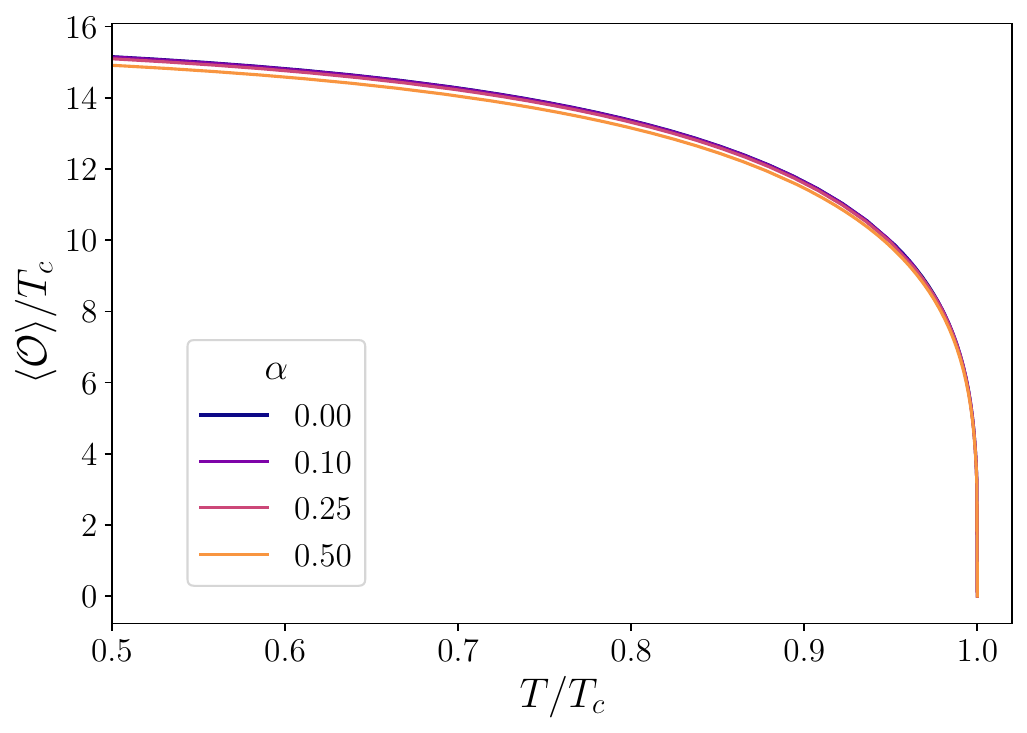}}
\subfloat{\includegraphics[width=0.5\linewidth]{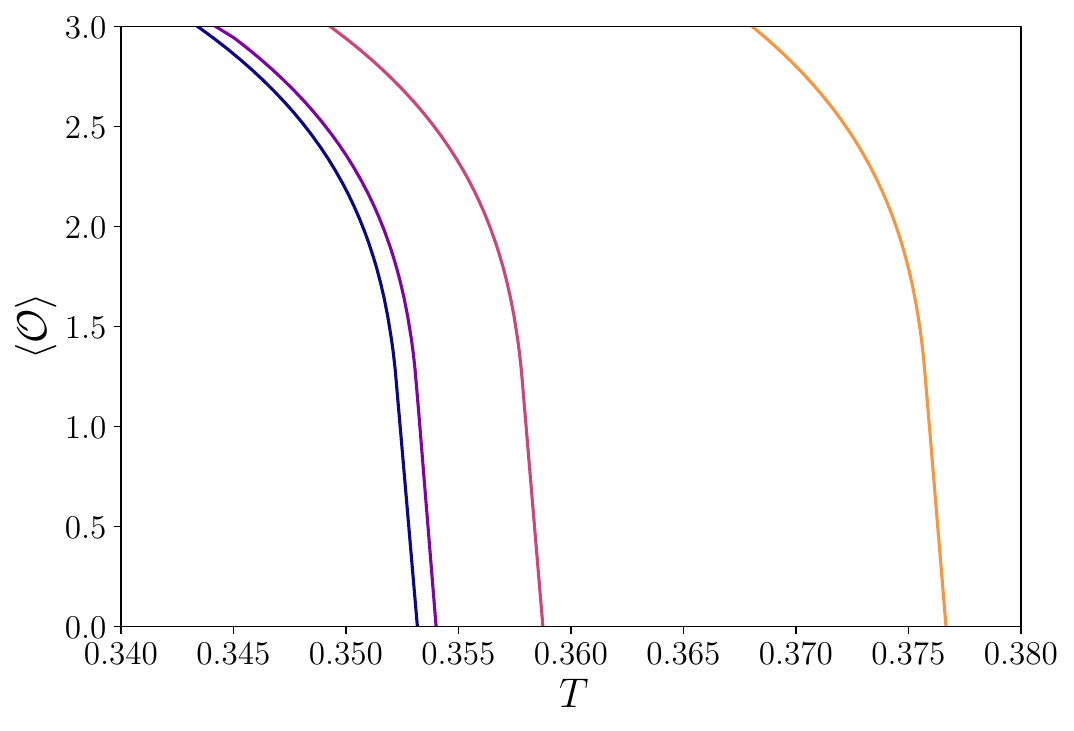}}
\caption{The superconductor-normal phase transition driven by the temperature, affected by {\it dark matter}. The right panel shows the zoom of the area of left plot which is close to the phase transition and without temperature normalisation. The coupling with the \textit{dark sector} causes the increase of critical temperature.}
\label{fig:temp_PT}
\end{figure}

\begin{figure}
\centering
\includegraphics[width=0.5\textwidth]{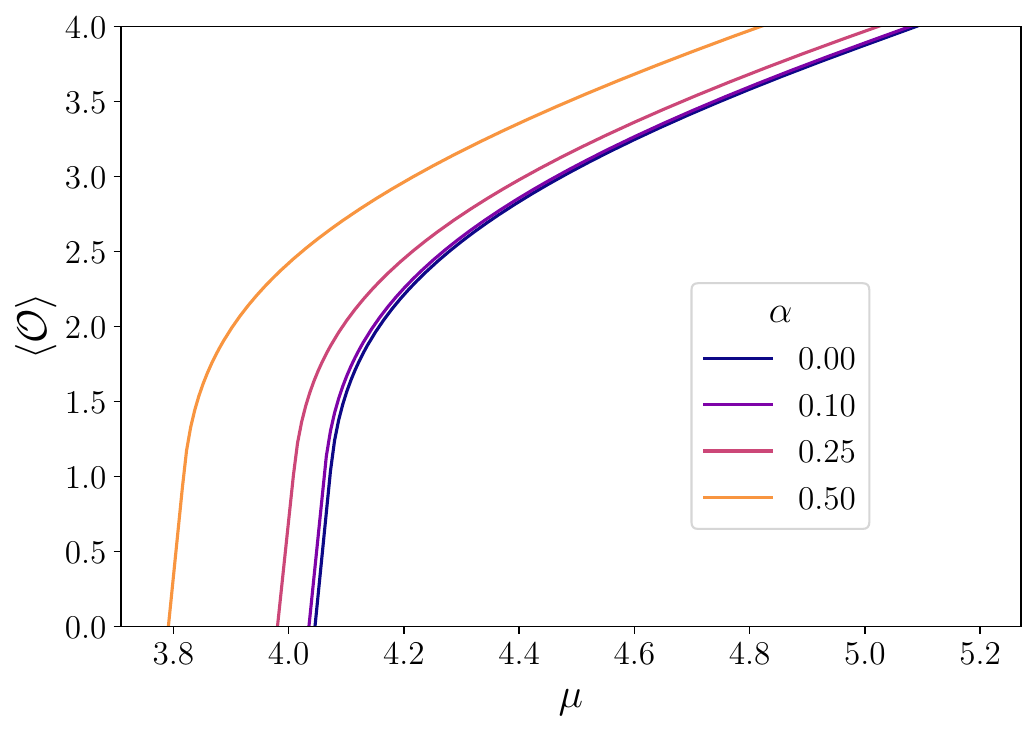}
\caption{The normal to superconducting phase transition driven by the change of the chemical potential (which can be interpreted as doping) for different values of
{\it dark matter} $\alpha$-coupling constants. The presence of the \textit{dark sector} coupling causes the phase transition for smaller value of the chemical potential.}
\label{fig:mju_PT}
\end{figure}

These equations establish the boundary conditions for a homogeneous region of superconductor in $x$-dependent approach, 
but for the analysis of phase transitions we set the current to zero, which implies that $M_x = 0$.
Now we can move towards numerical solution of our equations. In order to proceed further, 
we perform coordinate transformation $z = 1 - \frac{1}{r}$ in order
to work on a $(0,1)$ grid, where $z = 0$ is the black hole horizon and $z=1$ is the holographic boundary.
We solve the boundary value problem 
using the standard relaxation method for different values of temperature and chemical potential, to investigate the superconductor-normal phase transition. 
In figures \ref{fig:temp_PT} and \ref{fig:mju_PT} we plot the condensation operator $\langle O \rangle$ as a function of temperature and the chemical potential 
for the different values of {\it dark matter} 
$\alpha$-coupling constants.

One concludes that the bigger $T/T_c$ is the smaller value of $\langle O \rangle/T_c$ we obtain. In the case when we fix the value of $T/T_c$, then the bigger $\alpha$-coupling constant is taken into account the smaller $\langle O \rangle/T_c$ we get. Just, the condensation diminishes when {\it dark matter} coupling constant grows.
On the other hand, the bigger $\mu$ one considers, the larger value of $\langle O \rangle$ we receive, for the fixed value of $\alpha$-coupling constant. In the case when
$\alpha$ grows, we deduce that, the larger $\alpha$ is, the smaller value of the condensation operator one perceives.

In the both types of the analysed phase transitions, the presence of {\it dark matter} does not shift the phase transition point (critical temperature, critical chemical potential), yet it changes the value of condensation.

\begin{figure}
\centering
\includegraphics[width=0.5\textwidth]{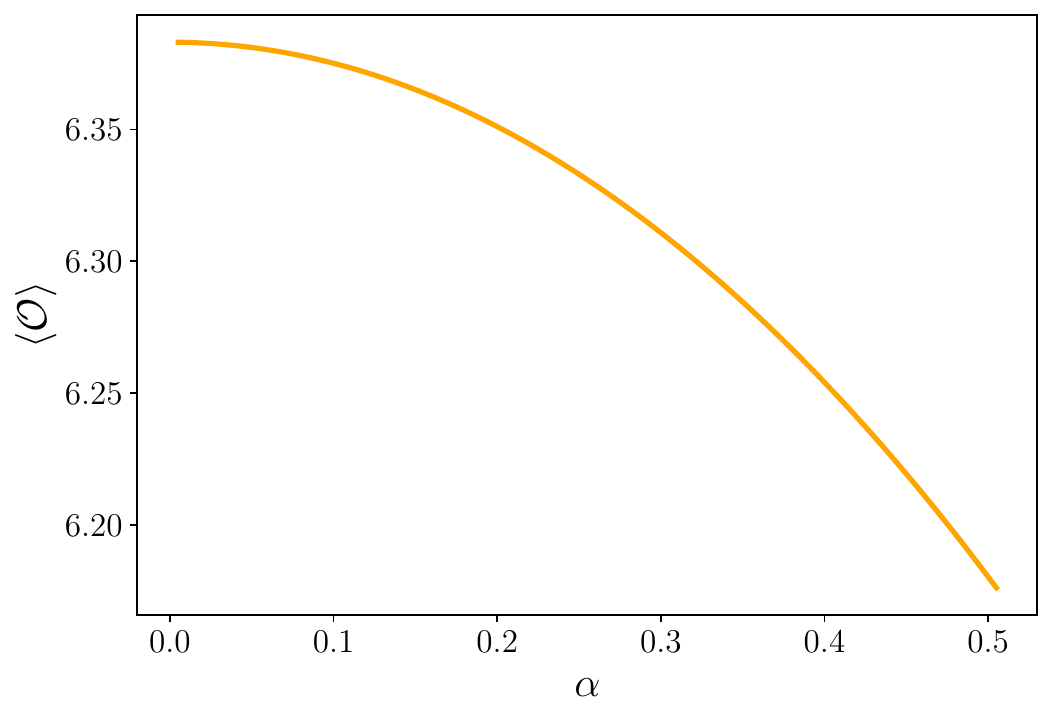}
\caption{Value of the condensation operator for the constant value of the chemical potential $\mu = 6$ and temperature $T = 3/4\pi$ versus the {\it dark matter} coupling
constant.}
\label{fig:cond_alpha}
\end{figure}

Figure \ref{fig:cond_alpha} depicts the relationship between the condensation operator and {\it dark matter} coupling constant, for the fixed temperature and chemical potential.
The differences are not too big, but they are visible and the lines are distinguishable.
However there might be a problem with measuring such a deviation. 
Our idea to overcome this obstacle is presented in the next section -- by using superconducting devices.

The calculations of this section served as a test of our numerical procedure. The procedure 
passed the test as the closer inspection of the equations of motion (\ref{eq:hom1}) - (\ref{eq:hom3}) 
shows that the parameter $\talpha$ can be eliminated by rescaling the $\psi$ field, as $\tilde{\psi}={\psi \over \sqrt{\talpha} }$,
with $\tilde{\psi}$ being independent on $\alpha$. This rescaling leads to the similar rescaling of the condensation 
parameter $\langle O \rangle$, leading to the solution with $\alpha$ independent value $\langle O \rangle_0$. The
expected behavior  $\langle O \rangle=\sqrt{\talpha} \langle O \rangle_0$ is clearly visible 
in figure \ref{fig:cond_alpha} as are the concomitant changes of other parameters depending on $\langle O \rangle$.

\subsection{Analysis of the holographic SQUID}

In this subsection we shall treat the case when one has the strength function $\zeta$ controlling the interaction between {\it visible} and {\it hidden} sectors.

\begin{figure}
\centering
\includegraphics[width=0.5\linewidth]{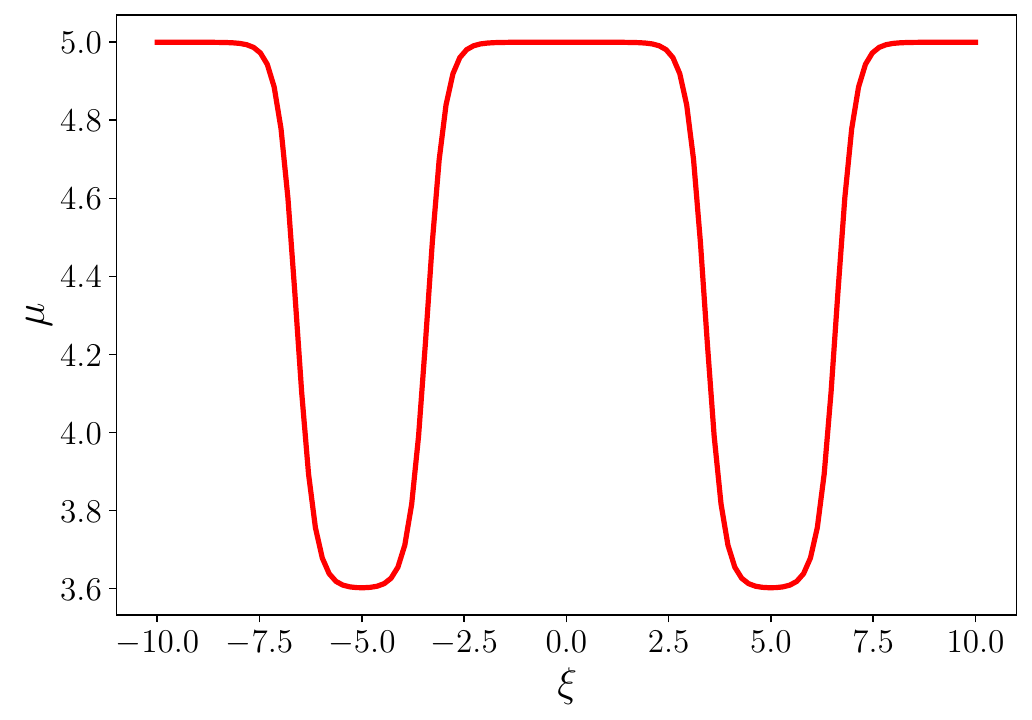}
\caption{The schematic profile of the chemical potential in the superconducting quantum interference 
device consisting of two holographic Josephson junctions (lower values of $\mu$) embedded into the superconducting loop 
with $\mu=5$. }
\label{fig2p}
\end{figure}

To imitate the two insulating regions embedded into the superconducting ring we take the
chemical potential as given in \cite{caisq14}
\be
\mu(\xi) = h - \sum_{i=1,~2} d_i~\bigg[
\tanh \bigg( \frac{k_i~(\xi-p_i+ w_i)}{\pi} \bigg) - \tanh\bigg( \frac{k_i~(\xi-p_i - w_i)}{\pi} \bigg) \bigg],
\label{mu-shape}
\ee
where i = 1, 2 stand for two junctions in the SQUID ring, and $h,~d_i,~k_i,~p_i$,
and $w_i$ are related to the highest value of the chemical potential (inside the
superconductor), depth, slope, position, and width of the junction $i$, respectively.
The idea behind this choice is the following:
from the previous studies we know that with and/or without {\it dark matter} it is the value of
the chemical potential which tunes the transition between insulator and metal and
the temperature decrease induces metal - superconductor transition. To construct 
a junction it is thus enough to make  the chemical potential in the normal part
of the junction lower than its critical value for superconducting transition, 
while leaving it above this value in the superconducting parts of the junction. 
This is just embodied in the equation (\ref{mu-shape}).
and shown in figure \ref{fig2p}, for
the weak links located at $p_1=5$ and $p_2=-5$ with $h=5$ and identical other parameters 
$d_1=d_2=0.7$ widths $w_1=w_2=1.6$ and slopes $k_1=k_2=7$.

In the discussion of the holographic SQUID we change 
$\ga_L$ and $\ga_R$, defined in what follows.
In our model we assume that the {\it dark matter} is present only in one of the junctions of SQUID. 
 Such approach gives us a possibility to calculate the difference of the phases between junctions caused by the {\it dark matter} coupling.
In the trivial case (without the presence of {\it dark matter} sector) the effective
magnetic flux defined as $\Phi = \gamma_L - \gamma_R$ vanishes, therefore the critical current
\begin{equation}
J_c = \sqrt{J_{L}^2 + J_{R}^2 + 2 J_{L} J_{R} \cos(\Phi)},
\label{eqn:squid_Jc}
\end{equation}
becomes constant and equals to $|J_{L} + J_{R}|$
On the contrary, when {\it dark sector} particle is present 
in one of the junctions, we receive a "yes or no" type criteria for the {\it dark matter}  occurrence.

In adverse to the previously obtained results \cite{caisq14},
we do not impose the continuity conditions on our fields in the node points $\xi = 0$ and $\xi = \pm 10$.
Moreover we argue that these points can be singular due to the presence of the additional source terms. Especially the supercurrent might not be well behaved when it inflows and outflows from our system.
If we wish to model the SQUID with two currents flowing parallel, we have to take $J \sim sgn(\xi)$, 
because current flows into the system at $\xi = 0$, then in one branch it flows to $\xi = - 10$ 
and in the another to $\xi = 10$, against and with the $\xi$ axis, respectively.

As far as the 
boundary value problem is concerned, for the AdS/CFT boundary one has the asymptotic expansions, for the black hole horizon we require that $M_t = 0$, while the conditions for the remaining functions are given by the adequate equation of motion.
On the other hand,
for the $\xi$ boundary, in both cases,  we impose that the functions $\vert \psi \vert$, $M_t$ and $M_x$ have to be even and $M_r$ is an odd function with respect to the central point of the junction.

\begin{figure}
\centering
\includegraphics[width=1.\textwidth]{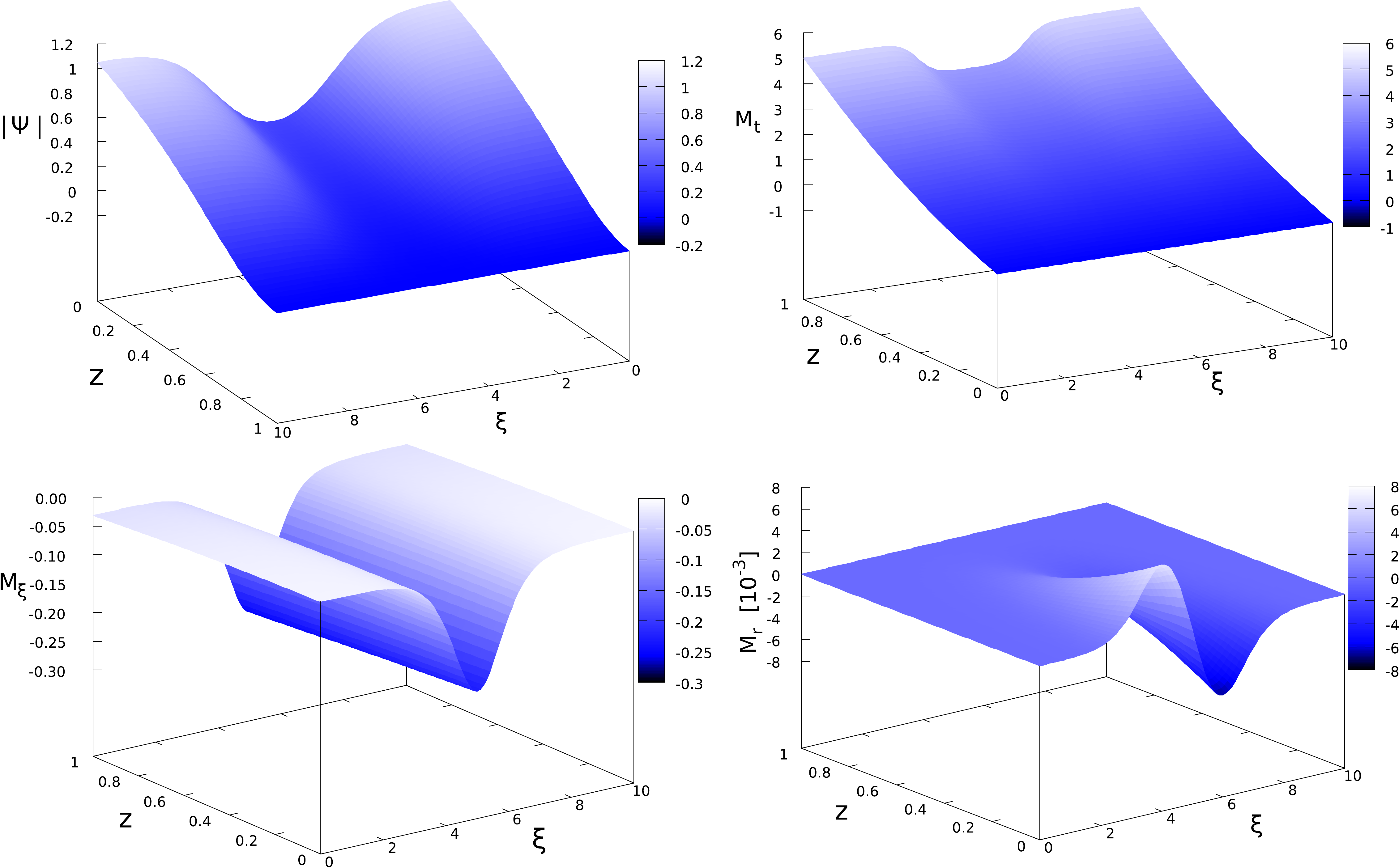}
\caption{Plots of numerical solutions of equations of motion (\ref{eq:systemstart})-(\ref{eq:systemend}) for one of holographic SQUID's Josephson junction with parameters $J = 0.03$ and $\alpha_0 = 0.2$. We can see the imposed chemical potential boundary condition for $M_t$ at the AdS boundary and the reaction of the remaining functions to it.}
\label{eqs}
\end{figure}

In our numerical computations we implement the pseudo-spectral method with Chebyshev
points, which constitutes an extremely efficient tool for smooth functions we deal with in the problem.
For convenience we compactify the $r$-coordinate using following transformation $z = 1 - 1/r$, after which $z = 0$ corresponds to the black hole horizon and $z = 1$ to the AdS boundary. Also we squeeze $\xi$ dimension into $(-1,~1)$ range which is the domain of the Chebyshev polynomials.
Generally pseudo-spectral methods require only few points in the spatial discretisation to achieve satisfying results, therefore we use the grid with 37 points along $\xi$ direction and 20 points along $z$ axis.
By expanding the functions into Chebyshev series we translate a differential problem stated by equations of motion into an algebraic one. Then we deal with a system of nonlinear algebraic equations by standard Newton-Raphson method.
In the explained way we solve numerically our set of partial differential equations (\ref{eq:systemstart})-(\ref{eq:systemend}) for different values of $J_0$, varying from $-0.06$ to $0.06$. 
In figure \ref{eqs} we plot the exemplary solutions of the underlying equations of motion for $J = 0.03$ and $\alpha_0 = 0.2$.
We can see the imposed shape of the chemical potential at the boundary $z=1$ of the $M_t$ function. The remaining functions react accordingly to the presence of the normal phase. The shapes of the solution are similar to the ones obtained in \cite{hor11,wan12,caisq14,tak15}. However it contains the traces of the influence of $\alpha$ coupling, but they are too small to be visible on such 3D plots.
Nevertheless they do have the influence on some of the properties, which will be elaborated below.

After solving our differential equations we obtain the phase differences for the left and right 
Josephson junctions separately, using the following formula:
\begin{equation}
\gamma_L = -\int_{-10}^0 \Big[ \nu(\xi) - \nu(0) \Big] d\xi,
\end{equation}
\begin{equation}
\gamma_R = - \int_{0}^{10} \Big[ \nu(\xi) - \nu(0) \Big]d\xi.
\end{equation}
Next, for each of the Josephson junction we fit in a sine relation between $J_{L/R}$ and $\ga_{L/R}$, to obtain the value of the maximum current.

\begin{figure}
\centering
\includegraphics[width=0.7\textwidth]{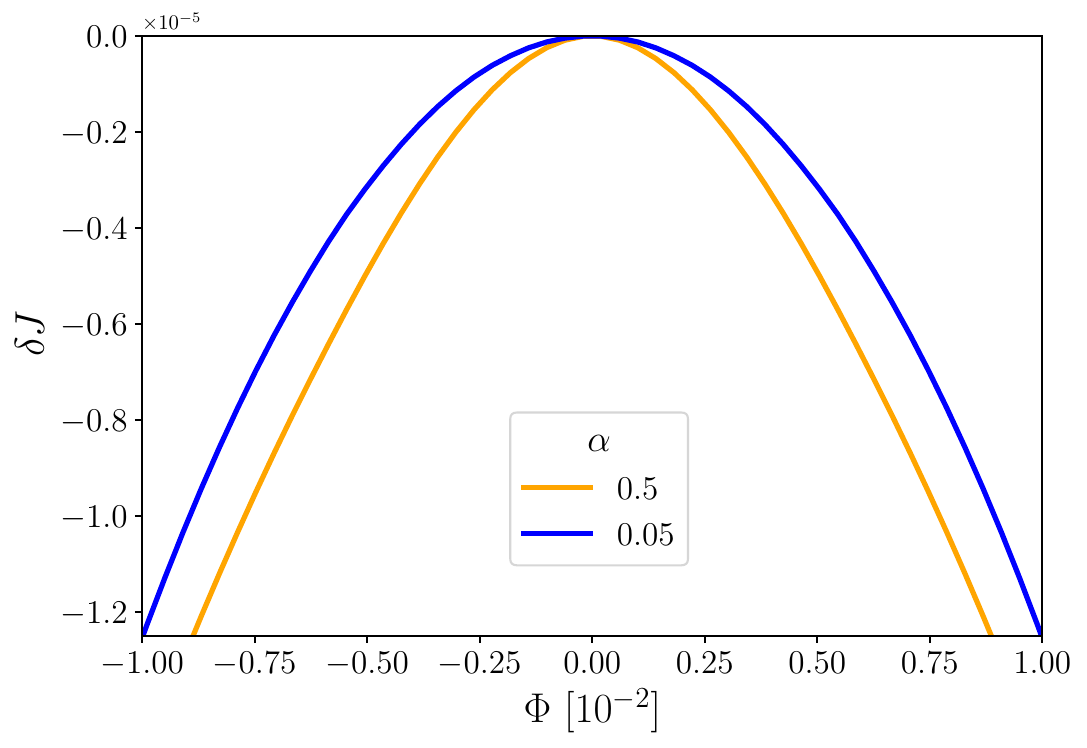}
\caption{Relative change of critical current ratio from the relation (\ref{eqn:squid_jc_ratio}) versus 
the effective magnetic flux, which appears to be non-zero in the presence of 
{\it dark matter}. The yellow curve represents solution when $\alpha_0 = 0.5$, while the blue one $\alpha_0 = 0.05$. In the case with $\alpha = 0$ such a dependence would not occur, 
resulting a trivial SQUID with two identical Josephson junctions and $\delta J\equiv 0$.
}
\label{fig:squid_jc_flux}
\end{figure}

Having obtained the values of the critical currents for each junction we can proceed to calculation 
of the {\it dark matter} particle induced effective magnetic flux $\Phi$, as defined previously.
We wish to see how the critical current of holographic SQUID, defined by the relation  (\ref{eqn:squid_Jc}), 
changes with the presence of {\it dark matter} in the system, hence one elaborates the case when
$\zeta \neq 0$.
To proceed further let us define the critical current ratio, given by
\begin{equation}
\delta J = \frac{J_c(\Phi) - J_c(0)}{J_c(0)}.
\label{eqn:squid_jc_ratio}
\end{equation}
In figure \ref{fig:squid_jc_flux} the critical current ratio as a function of the magnetic flux, for 
different values of $\alpha_0$, is depicted.
In our model such behaviour is only possible when {\it dark particle} appears in one 
of the junctions. Otherwise, if $J_c$ is independent of $\Phi$ the interference does not occur.
This gives us the possibility to establish the criterion for detection of the discussed coupling 
which is correlated with the {\it dark sector}.

While the residuals of numerical solution of our equations (\ref{eq:systemstart})-(\ref{eq:systemend}) and the constraint relation (\ref{eq:constraint}) are relaxed to the values less than $10^{-12}$, the relation from the figure \ref{fig:squid_jc_flux} carries a burden of inaccuracy of the least squares method. Namely the sine function is not a perfectly fitting one and the best result we can obtain for the norm of residuals is to the order of $10^{-6}$. It results that
the uncertainty of the critical current is $\sim 10^{-4}$.
Having this in mind, for little and yet realistic values of coupling our computed values of $J(\Phi)$ are below the level of accuracy, so we cannot state its direct value. Although when working on differences, like in our ratio given by the relation (\ref{eqn:squid_jc_ratio}), these uncertainties are subtracted, so using this quantity seems sensible.

\begin{figure}
\centering
\includegraphics[width=0.7\textwidth]{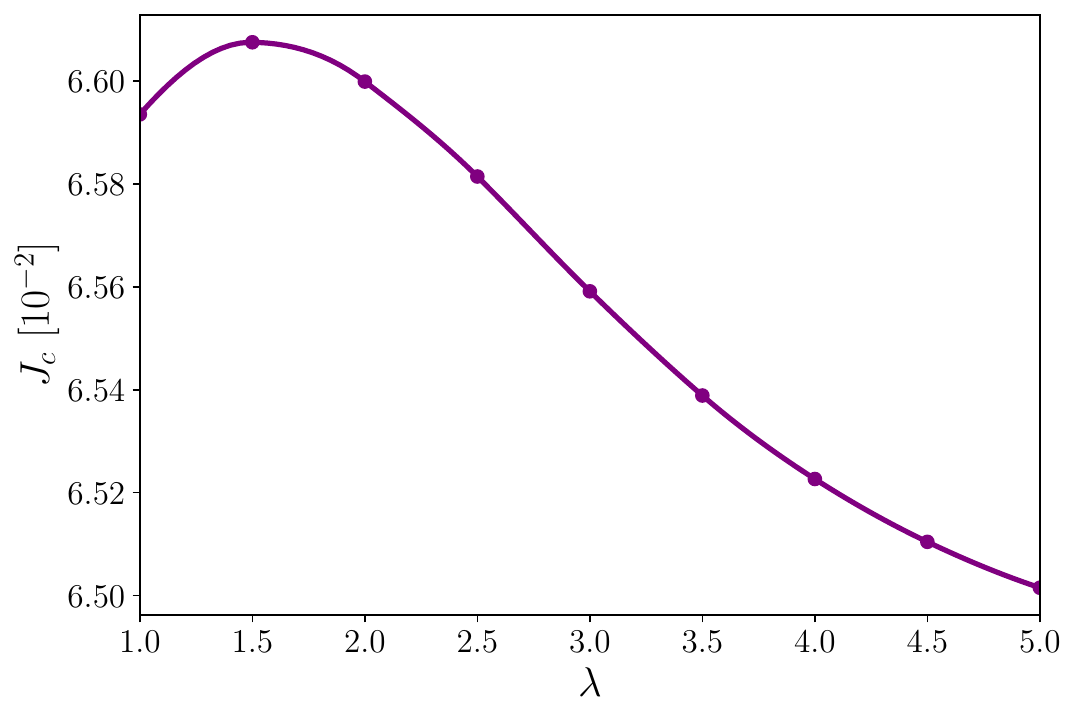}
\caption{The critical current of {\it dark matter} influenced Josephson junction 
as a function of Gaussian broadening $\lambda$ from the relation
(\ref{eqn:gauss_alpha}). It is clearly seen that there is a maximal sensitivity for some value of $\lambda$. 
In this plot we used $\alpha_0 = 0.5$, the size of the junctions is 1.6, while the separation between the junctions is 10, in the units used in the plot.}
\label{fig:jc_lambda}
\end{figure}

We also investigated the influence of increasing the Gaussian packet decay length $\lambda$ 
on the current-phase relation of one of the holographic SQUID's Josephson junctions.
By implementing to our code various values of $\la$ (from 1 to 5), 
we have found out an interesting behaviour which is depicted in figure \ref{fig:jc_lambda}.
It envisages the critical Josephson current as a function of the wavelength of the decay length $\la$.
This plot might be interpreted as the sensitivity of the holographic SQUID on detecting {\it dark matter} sector. 
On the other hand, it is a very interesting result and may constitute a guidance for future experiments.
Namely, by adjusting the size of the normal/insulating part of the holographic Josephson junctions 
one can tune up the system to react most sensitively on possible detection of {\it dark particle}.
The obtained value of the optimal $\la\approx 1.5$ is correlated, as it is expected to be,
with the width $w$ 
of the normal region of the junction affected by the {\it hidden particles}.

Having a possible range of the Gaussian packet decay length $\lambda$,
one receives a possibility 
to prepare an adequate detector for measurements. In the present approach we are assuming that the {\it dark matter}
particle passes the center of the junction. However, one expects that {\it dark particles} move slightly
off the center. In such situation the sensitivity can be roughly read of from the curve shown 
in figure \ref{fig:jc_lambda}, as the decrease/increase 
of $\la$, mimicking the distance 
from the center of the holographic junction.  


\section{Holographic versus condensed matter SQUID}
Here we compare the holographic results with the expectations based on the 
standard theory of laboratory  SQUID as shown in figure \ref{fig2}.  
Each branch of the SQUID, in figure \ref{fig2}, contains one Josephson junction.
In general, if the device is biased  the current phase relation reads
\be
I=I_{max}\sin\left(\Delta \phi+\frac{2eV}{\hbar}t\right),
\label{jos-gen}
\ee
where $V=(\mu(\infty)-\mu(-\infty))/e$ is the voltage drop across the junction, $\Delta \phi$ is earlier introduced phase difference between two superconductors. Equation (\ref{jos-gen}) shows that the current of the biased  junction  oscillates  with frequency $\frac{2eV}{\hbar}$. In the following, in full analogy to the holographic treatment we shall
consider only direct current (dc) SQUID {\it i.e.}  with $V=0$.

The presence of the magnetic flux perpendicular to the plane of the device induces the screening currents in the superconductor and changes the phase relations in both branches. By the gauge invariance principle one finds that the flux $\Phi$ penetrating the SQUID loop modifies the relation between the phases of two junctions A and B and one gets \cite{annett}  
\be
\Phi=2\pi\phi_0(\Delta \phi_A-\Delta \phi_B).
\label{flux}
\ee
The total current $I$ entering the device from the left splits into $I_A$ and $I_B$.
At the right hand side of the device both $I_A$ and $I_B$ fulfilling the relation (\ref{eqjj})
add to give again the current $I$. One thus gets (for $I_A=I_B$ and with the phase 
differences $\Delta \phi_A$ and $\Delta \phi_B$ in the respective branches)
\be
I=I_A+I_B=2I_{max}\sin\bigg(\frac{\Delta \phi_A+\Delta \phi_B}{2}\bigg){\cos\bigg(\frac{\Delta \phi_A-\Delta \phi_B}{2}\bigg)}.
\ee
As was shown in  \cite{ketterson1999} the phase difference is related
to the total flux $\phi_{total}$, which for the negligible inductance of the loop reduces
to the external flux $\Phi$ as in the equation (\ref{flux}). With the symmetric distribution 
of the phase modifications
\ben
\Delta \phi_A=\Delta\phi+\frac{\pi\Phi}{\phi_0} \\
\Delta \phi_B=\Delta\phi-\frac{\pi\Phi}{\phi_0}
\een
 one gets
\be
I=2I_{max}\cos\bigg(\frac{\pi\Phi}{\phi_0}\bigg)\sin\bigg(\Delta\phi_B+\frac{\pi\Phi}{\phi_0}\bigg)
=2I_{max}\sin(\Delta \phi)\cos\bigg(\frac{\pi\Phi}{\phi_0}\bigg),
\ee
where $\phi_0\equiv hc/2e\approx 2.07\cdot 10^{-7}Gs\cdot cm^2$ is the quantum flux.
One sees that the amplitude of the current in the SQUID is modulated by the factor 
$|\cos(\frac{\pi\Phi}{\phi_0})|$ and the ratio of the maximal currents flowing in the device subject to the external flux to that without it is given by this factor
\be 
\frac{I_m(\Phi)}{I_m(0)}=\left|\cos(\frac{\pi\Phi}{\phi_0})\right|.
\label{Bperp}
\ee

\begin{figure}
\centering
\includegraphics[width=0.7\linewidth]{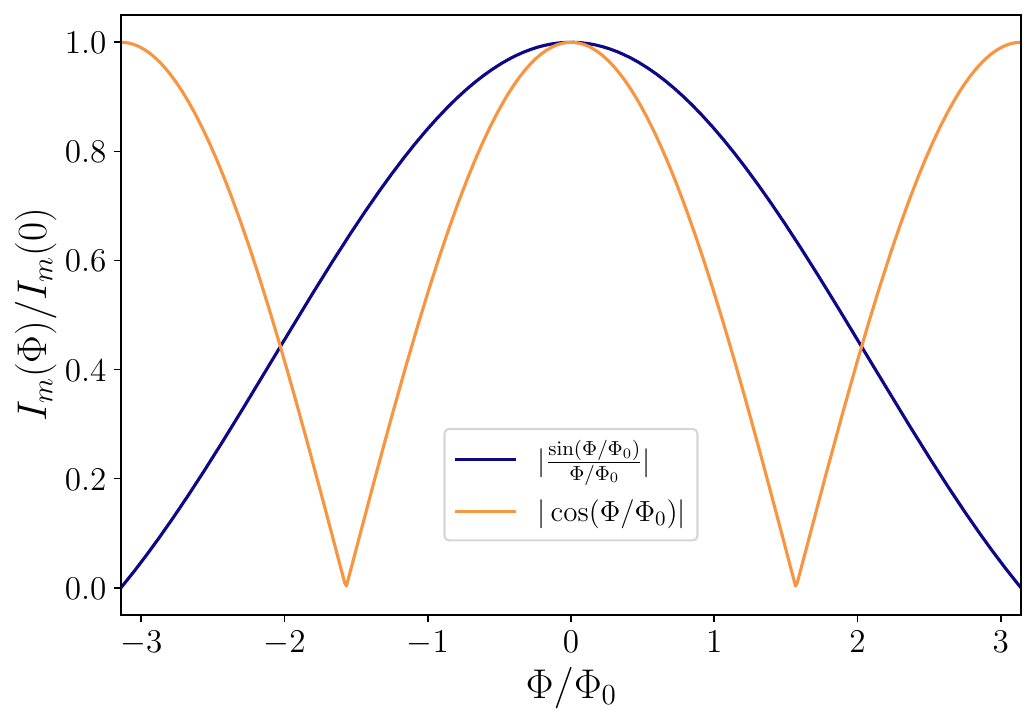}
\includegraphics[width=0.45\linewidth]{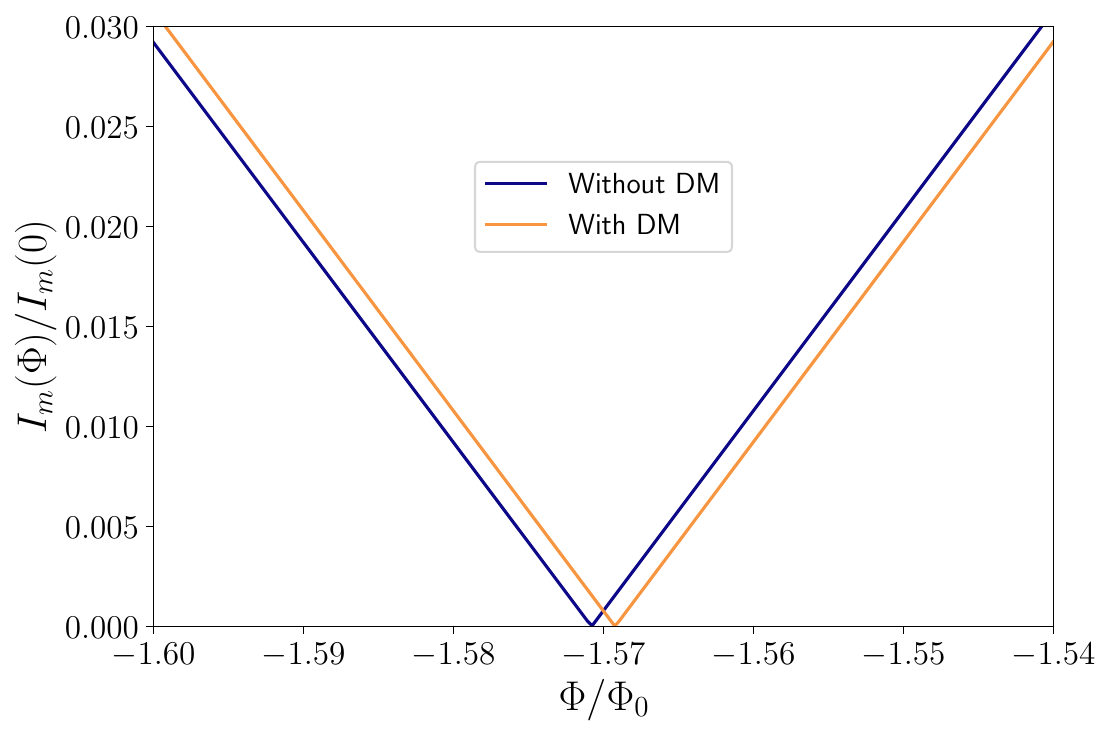}
\includegraphics[width=0.45\linewidth]{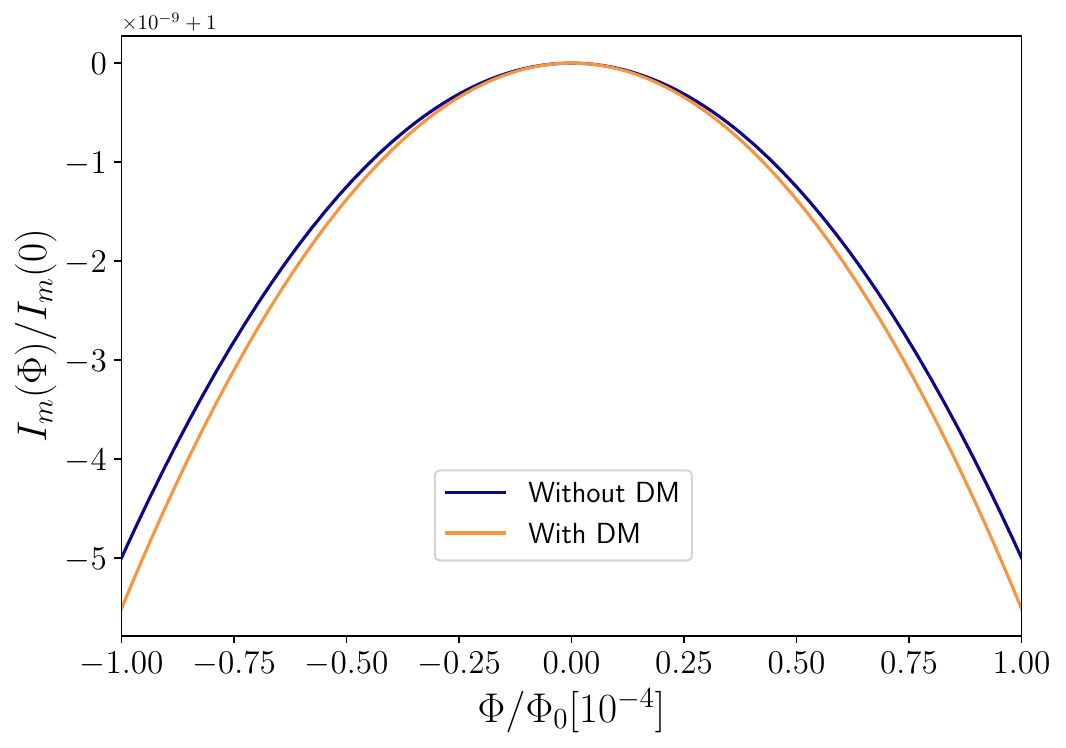}
\caption{The interference patterns in SQUID is shown in the upper panel. Left-lower
panel shows the detail SQUID responses for magnetic field tuned close to the 
minimum. The passage of {\it dark particle} shifts the minimum and the SQUID responds with a well visible signal (lower left panel $\alpha=0.004$). The detection of the signal is much more complicated if the SQUID is tuned close to the maximum of the current (lower right panel). The differences are small even for very large values of the coupling, which in this panel is assumed to be $\alpha=0.2$. }
\label{fig:j-j-interf}
\end{figure}

For the completeness  it has to be noted that if the magnetic flux $\Phi$ is piercing one of the weak links (junctions), then another ``diffraction pattern'' is observed \cite{tinkham}. It may be characterised by  the ratio of the maximum current in the device with ($I_m(\Phi)$) and without ($I_m(0)$) extra flux
\be
\frac{I_m(\Phi)}{I_m(0)}=\bigg|\frac{\sin(\frac{\pi\Phi}{\phi_0})}{\frac{\pi\Phi}{\phi_0}}\bigg|.
\label{Bparalel}
\ee
The above two possibilities to change the maximum current in the device formulated by equations (\ref{Bperp}) and (\ref{Bparalel}) by applying to it the external magnetic field are plotted in the upper panel of the figure (\ref{fig:j-j-interf}).
The blue curve on this panel shows the current as a function of the magnetic flux ratio $\Phi/\phi_0$ for the flux penetrating the extended junction and the magnetic field parallel to the plane of the SQUID, while the other curve shows standard geometry with the flux penetrating the loop. 


Now we discuss the possible effect of {\i dark matter} particle on the current through the condensed matter SQUID. For this we imagine the {\it dark matter particle} traversing one of its junctions. In order to find the changes of the flux $\Phi$ induced by the passage of the {\it dark particle}, let us consider the action (\ref{sgrav}), and define new gauge fields which enable us to get rid of the 
{\it kinetic mixing } term. Namely one has
\ben
\tA_\mu &=& \frac{\sqrt{2 -\alpha}}{2} \Big( A_\mu - B_\mu \Big),\\
\tB_\mu &=& \frac{\sqrt{2 + \alpha}}{2} \Big( A_\mu + B_\mu \Big).
\een
It leads to
\be
\frac{1}{4} F_{\mu \nu} F^{\mu \nu} +
\frac{1}{4}B_{\mu \nu} B^{\mu \nu} + \frac{\alpha}{4} F_{\mu \nu} B^{\mu \nu}
\Longrightarrow
\frac{1}{4} \tF_{\mu \nu} \tF^{\mu \nu} +
\frac{1}{4}\tB_{\mu \nu} \tB^{\mu \nu},
\ee
where $\tF_{\mu \nu} = 2 \na_{(\mu }\tA_{\nu )}$ and respectively $\tB_{\mu \nu} = 2 \na_{(\mu }\tB_{\nu )}$. The corresponding magnetic fluxes are provided by 
\ben
\Phi_{\tA_\mu} &=& \frac{\sqrt{2 - \alpha}}{2} \Bigg[ \frac{2 \pi ~a}{e} - \frac{2 \pi~b}{e_d} \Bigg],\\
\Phi_{\tB_\mu} &=& \frac{\sqrt{2 + \alpha}}{2} \Bigg[ \frac{2 \pi ~a}{e} + \frac{2 \pi~b}{e_d} \Bigg],
\een
where $e$ stands for the charge of Maxwell field while $e_d$ is connected with the charge of {\it dark matter} sector gauge field, while $a,~b \in {\bf Z}$.

Thus the total magnetic flux encompassed  by the SQUID is given as $\Phi_{\tA_\mu} + \Phi_{\tB_\mu} $. 
 Having in mind that the coupling constant $\alpha \ll 1$, one can rewrite the sum as follows:
\be
\Phi_{\tA_\mu} + \Phi_{\tB_\mu}  \simeq 
\frac{2 \sqrt{2} \pi}{e} \Bigg[ a + \frac{\alpha}{4} \Big( \frac{e}{e_d} \Big) ~b\Bigg].
\ee
If we assume that $a=b$ it leads to the relation
\be 
\Phi_{\tA_\mu} + \Phi_{\tB_\mu}  \simeq  \Phi_A \Bigg[ 1 + \frac{\alpha}{4} \Big( \frac{e}{e_d} \Big) \Bigg],
\label{eff-fux}
\ee
where $\Phi_A$ is the magnetic flux bounded with Maxwell field. \\

 On the basis of the equation (\ref{eff-fux}) and assuming that $\frac{e}{e_d} = 1$ we argue that due to the kinetic mixing (with the coupling $\alpha$) the effective magnetic field and thus the flux through the device would  change from the value $\Phi$ to the new value $(1+\alpha/4)\Phi$ when the {\it dark particle} is traversing the SQUID, so obviously the current in the device would also change. This change is illustrated in the lower panels of figure (\ref{fig:j-j-interf}).

 The lower left panel of figure (\ref{fig:j-j-interf}) compares the detected signals close to its minimum. The blue curve is the signal without {\it dark particle}. The zero current is expected at $\Phi/\phi_0=\pi/2$. If the {\it dark particle} appears in the SQUID it changes the flux across the device. This change of the flux obtained for $\alpha=0.004$ is visible as a nonzero current at the same external magnetic field value. In the figure both curves have been plotted as a function of the external flux $\Phi$ (no {\it dark particle} in the SQUID), which value is changed (locally in space and time) to $(1+\alpha/4)\Phi$ (when {\it dark particle} traverses the device).
 
On the contrary if one looks at the SQUID signal close to the maximum current (lower right panel) the effect is hardly visible for the same value of $\alpha$ (not shown). In the panel we are comparing two signals using $\alpha=0.2$. This large value of the coupling makes the signal visible. However, even the tiny change of the effective flux causes the shift of the minimum in the diffraction curves and makes the tiny effect visible. Thus the response of the SQUID which is tuned by the external \textbf{B} field to the minimum of the current seem easier to be detected. This is due to the fact that in the latter case one expects the sudden increase of the current from zero to some finite (albeit small) value. Unless the changes of the current are below the thermal or quantum fluctuations in the system the effect should be observable. This depends on the value of the kinetic mixing $\alpha$ and thus on the mass(es) of the {\it dark particle(s)}.  To define real experimental conditions one should also take into account fluctuations of the external flux. The field theory analysis of the possibility of detecting {\it dark matter} particles is an important and complicated subject~\cite{hochberg2016}, which we shall not pursue here any further. 

\section{Discussion and conclusion}
Using the holographic approach we have studied the influence of  the {\it dark sector} particle on the  the holographic model of the SQUID. To this end we have assumed gravity background 
consisting of the four-dimensional  static AdS Schwarzschild black brane.
The holographic modelling of the device is achieved by use of the special form of the chemical potential mimicking the two insulating/normal regions embedded into superconducting ring. Moreover one assumes that {\it dark matter} sector is present only in one of the holographic Josephson junctions of the SQUID under inspection.

This supposition enables us to calculate the difference of the phases between the junctions, caused by {\it dark matter particle}.
We solve numerically equations of motion and receive values of $M_i$ and thus the phase, 
for the specific  choice of the current $J$ and $\zeta$ strength parameter function, controlling the interaction of {\it visible} and {\it hidden} sectors. 
This procedure leads to the critical current and the phase for each of the examined holographic junctions and allows establishing 
the current phase relation which has a typical sinusoidal character.

The passage of the {\it dark sector particle} by one of the holographic SQUID junctions modifies the phase difference in that junction and leads to the modification of the critical current. This shows up as the appearance of the {\it dark sector}  induced effective flux $\Phi$. The difference of the critical currents of the SQUID measured for $\zeta\ne 0$ and $\zeta=0$, signals the {\it dark sector} field detection.

We have assumed that the presence in the junction of the {\it dark matter} sector field can be described by the 
Gaussian type of dependence of the coupling $\zeta$ on the $\xi$ coordinate. It turns out  that the  
increase of  the Gaussian packet decay length $\lambda$ has an effect on  the current-phase relation of one of the holographic SQUID's Josephson junctions.
Moreover, the critical Josephson current is a function  of the decay length $\la$.
This fact supplies us the tool for changing the sensitivity of the holographic SQUID in the dependence on the expected properties {\it dark matter} sector. 
Interestingly
the simplest possible  assumption that due to the kinetic mixing between the fields the {\it dark particle} traversing the condensed matter SQUID increases the effective magnetic field leads to the results similar to those obtained by means of holographic analogy.

To conclude, we remark that the presented results have been  obtained for the holographic s-wave Josephson junction and SQUID composed of such junctions. They envisage some features which can be utilised in future experiments aimed at detecting {\it dark matter} sector particles. Of course, we are aware that examination of more complicated models like p-wave or $p_x+ip_y$ might lead to even better detection methods.  We shall investigate these problems elsewhere.
\acknowledgments
We thank J.E. Santos and R. Moderski for valuable comments concerning the numerical methods. This work has been partially supported by the M. Curie-Sklodowska University and the  National Science Center grant DEC-2017/27/B/ST3/01911 (Poland).



\end{document}